\documentclass{PoS}
\usepackage{epsfig}
\usepackage{color}
\let\normalcolor\relax

\PoS{PoS(LAT2005)027}

\definecolor{darkgreen}{rgb}{0,0.5,0.5}
\newcommand{\blu}{\color{blue}}
\newcommand{\red}{\color{red}}

\newcommand{\dgre}{\color{darkgreen}}
\newcommand{\beq}{\begin{equation}}
\newcommand{\eeq}{\end{equation}}
\def\Tr{{\sf Tr}\,}

\title{Scalar glueball and meson spectroscopy
in unquenched lattice QCD with improved staggered quarks}

\ShortTitle{Scalar glueball and meson spectroscopy}

\author{Eric B. Gregory, \speaker{Alan C. Irving},
	Craig C. McNeile, Steven Miller, 
        Zbyszek Sroczynski\\
        Department of Mathematical Sciences,
        University of Liverpool\\
        Liverpool, U.K.\\
        E-mail: \email{aci@liv.ac.uk}}

\abstract{
We present results of an exploratory study of singlet scalar states
in unquenched QCD using both glueball and meson operators. Results for
non-singlet non-strange scalar mesons are also presented. We use Asqtad
improved staggered fermions and gauge configurations
generated by the MILC collaboration at lattice spacings of .12 and .09 fm.
In this formulation, the glueball mass is not significantly different from the
quenched value at finite lattice spacing. Significant taste violations
are present in the scalar sector. At light quark masses, decay channels
complicate the mass determinations. There is some evidence that
the non-strange singlet meson lies below the non-singlet meson.}

\FullConference{XXIIIrd International Symposium on Lattice Field Theory\\  
                25-30 July 2005\\
	Trinity College, Dublin}

\begin{document}

\section{The scalar sector of QCD}
The scalar glueball is by now relatively well studied in quenched QCD.
In full QCD with realistically light quarks it is 
not~\cite{Bali:2000vr,Hart:2001fp}
Simulations with
moderate to heavy Wilson-like quarks show
effects which may reflect poorly understood
lattice artifacts rather than continuum physics~\cite{Hart:2001fp}.
UKQCD plans to use its new QCDOC dedicated supercomputer to
create substantial data sets to complement the existing
MILC configurations~\cite{Bernard:2001av,Aubin:2004wf} 
which will allow substantially higher
statistics
and moderate to light quark mases.
Clearly, there are a number of serious challenges to be met.
On the one hand light quark masses help one probe mixing but at the same time
open up decay channels which complicate spectroscopy using current methods.
Furthermore, the singlet sector may be precisely that sector where the 
staggered quark method is at its most vulnerable because of the 
differences in the implementation of valence and sea quarks and because
of inadequately controlled taste-breaking effects in the sea.

\section{Glueballs}
\begin{figure}[htb]
\centerline{\epsfig{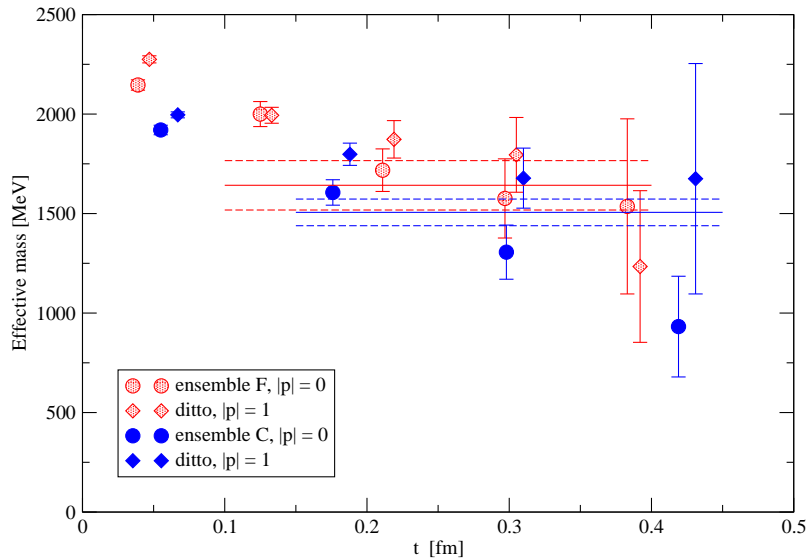}}
\caption{Effective masses using glueball operators.}
\label{fig1}
\end{figure}

\begin{figure}[htb]
\centerline{\epsfig{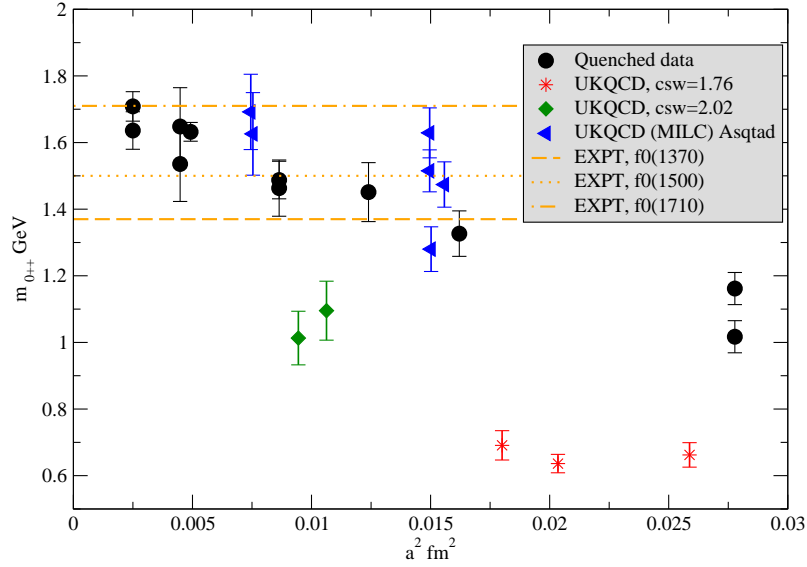}}
\caption{Compilation of glueball masses for quenched and unquenched QCD.}
\label{fig2}
\end{figure}

Figure~\ref{fig1} shows effective masses for ensembles $C$ and $F$
of the MILC data sets~\cite{Bernard:2001av,Aubin:2004wf} 
indicated in Table~\ref{tab1} below.
%
The comparison is made in physical units using measured values of $r_1/a$.
Non-zero momentum operators are also used to help confirm the mass estimates.
Horizontal lines indicate fitted values and errors. 
Figure~\ref{fig2} shows the results of such measurements superimposed on
a compilation of previous quenched and unquenched results.
We note that lattice artifacts appear smaller than for improved 
Wilson quarks with unimproved Wilson glue~\cite{Hart:2001fp}. 
The current errors are clearly not yet small enough to
resolve properly the experimental spectrum
or quark mass dependence.
\begin{table}[htb]
\begin{center}
\begin{tabular}{|c|c|c|c|c|c|}
\hline
Ensemble & $\beta$ & $am_{u,d}$ & $am_{s}$ & a [fm]& No. of configs \\
\hline
A & $6.81$ & $0.03$ & $0.05$ &  $0.1191$& $564$ \\
B & $6.79$ & $0.02$ & $0.05$ & $0.1196$ & $483$ \\
C & $6.76$ & $0.01$ & $0.05$ & $0.1215$ & $658$ \\
D & $6.76$ & $0.007$ & $0.05$ & $0.1209$ & $448$ \\
\hline
E & $7.11$ & $0.0124$ & $0.031$ & $0.0854$ & $514$ \\
F & $7.09$ & $0.0062$ & $0.031$ & $0.0860$ & $505$ \\
\hline
\end{tabular}
\end{center}
\caption{Main ensemble parameters}
\label{tab1}
\end{table}
%

\section{Scalar mesons}
A staggered operator which creates a state that lies
in the spin-taste representation $\Gamma_{S}\otimes \Gamma_{T}$
also couples to one lying in the $\gamma_{4} \gamma_{5} \Gamma_{S}
\otimes \gamma_{4}
\gamma_{5} \Gamma_{T}$ representation.
Thus a staggered meson correlator has the general form
\beq
{\cal C}(t) = \sum_n \bigl[A_n e^{-m_n(\Gamma_{S}\otimes
\Gamma_{T})t}
+ (-1)^{t} B_n
e^{-m_n(\gamma_{4} \gamma_{5} \Gamma_{S} \otimes \gamma_{4} \gamma_{5}
    \Gamma_{T})t}\bigr]
\label{eq:corr}
\eeq
In general, one therefore expects an oscillating contribution from a
parity partner of the desired state. For the scalar meson
($\Gamma_{S}\otimes \Gamma_{T}= 1\otimes 1$), the parity partner
is $\gamma_{4} \gamma_{5} \otimes \gamma_{4} \gamma_{5}$ which corresponds to
one of the pseudoscalars.
For the non-singlet scalar this is one
of the (taste-split) pions and so is a significant low lying
contribution which must be included in the fits.

For flavour singlet mesons, the correlator is of the form
\beq
{\cal C}(t) = C(t)-n_f D(t)
\label{eq:singlet}
\eeq
where $C(t)$ is the same correlator coupling to the NS meson state
and $D(t)$ is the disconnected (quark loops) correlator suitably
corrected ( by $1/4$) to account for the extra $4$ tastes that
can contribute as compared with the connected 
correlator~\cite{Venkataraman:1997xi}.
In this study $n_f=2$ and the $u,d$ quark masses are degenerate.
Note that there is no oscillating contribution to $D(t)$ in
(\ref{eq:corr}) since, in this case,
the parity partner would be taste non-singlet~\cite{Miller:2005lu}.

\section{Non-singlet meson}
\begin{figure}
\centerline{\epsfig{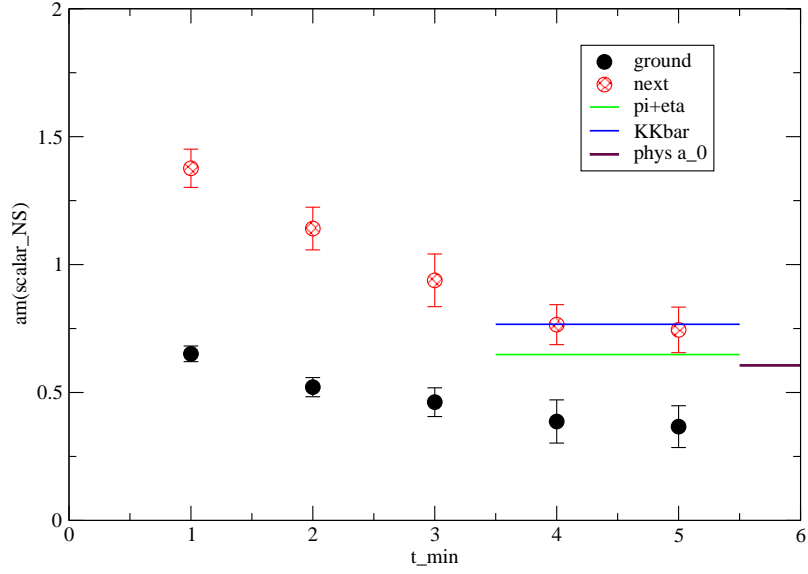}}
\caption{Non-singlet scalar mass as a function of minimum $t$ for
ensemble $C$.}
\label{fig3}
\end{figure}

Figure~\ref{fig3} illustrates (for ensemble $C$) the resulting
scalar masses from factorising fits
to eqn.~(\ref{eq:corr}) using local and fuzzed operators.
Three states were included - two scalar states and one oscillating pion.
The plot shows the effect of varying the minimum $t$-value used in
the fits. Here, $t_{\hbox{max}}$ was fixed at $11$ in lattice units.
%
The most noticeable feature is that
the \lq ground state\rq{} estimate
lies significantly {\em below} the effective
$m_\pi+m_\eta$ threshold expressed in lattice units.
Here the $\eta$ mass is estimated via the Gell-Mann Okubo mass formula
following~\cite{Bernard:2001av,Aubin:2004wf} and the pion
is taken to be the taste-singlet Goldstone pion.

The Figure also shows the $K\bar{K}$ threshold along with
the observed mass of the $a_0(980)$ expressed in lattice units.
This puzzling low-lying state was also observed by 
MILC authors~\cite{Aubin:2004wf}
using independent measuring techniques - a different choice of 
operators and correlators.

There are a number of other remarks to be made concerning these fits
and corresponding mass estimates.
For this ensemble (C - the lightest quark mass) the higher state appears 
roughly compatible (but possibly coincidentally) with the expected  
$a_0(980)$ mass and, hence with the nearby $K\bar{K}$ two-particle threshold 
estimated at this lattice spacing.

The oscillating parity partner state (pion) is well
determined and consistent with direct measurement using
a pseuoscalar operator. 
Qualitatively simliar behaviour is seen in the other ensembles ($A$ and $B$).
The fits did not require an excited oscillating state.
Fits where the lowest state was constrained to match the
expected mass of a $\pi+\eta$ state were unacceptable.

We recall that partially quenched studies of the NS scalar meson in other formulations
show abnormal correlator behaviour~\cite{Prelovsek:2004jp}. 
The present study could be regarded as partially quenched in the
sense that the valence and sea quarks are treated differently
with respect to taste - explicitly for valence operators but via
the fourth-root trick for the sea. 

Direct measurements of taste violations in the scalar NS sector are found
to be comparable with those found in the pion sector. The maximum
violation was of order $15\%$ on the coarse lattice ensemble $A$.
See~\cite{Miller:2005lu} for details.

Prelovsek, in these proceedings~\cite{Prelovsek:2005du}, 
has used
chiral perturbation theory for staggered fermions
to show how taste violations in this channel
can lead to effective G-parity violation in $a_0$ decay so that a small coupling
to $\pi\pi$ is allowed. This is then expected~\cite{Prelovsek:2005du}
to dominate the correlator at large 
Euclidean time and so could account for the behaviour seen in 
Figure~\ref{fig3}
and by MILC~\cite{Aubin:2004wf}.
 
\section{Singlet scalar with light quarks}

\begin{figure}
\centerline{\epsfig{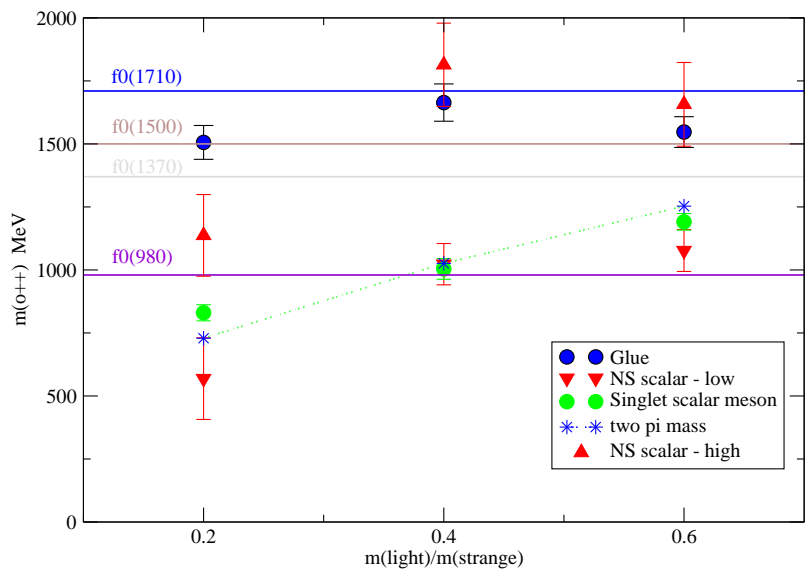}}
\caption{Singlet scalar mass as a function of quark mass.}
\label{fig4}
\end{figure}

The quark loops $L(t)=\Tr M^{-1}({\bf x},t)$ required for
$D(t)=<L^*(t)L(0)>_c$ were evaluated using
stochastic methods (with typicaly $48$ noise vectors).
Factorising fits, with local and fuzzed operators, were carried out
in which the oscillating term in $C(t)$ was fixed from the $NS$ fits
described earlier.
The singlet scalar mass extracted in this way is shown
as a function of the light quark mass in Figure~\ref{fig4}
({\dgre  green circles}).

%
%
For comparison we show again the masses obtained with
glueball operators along with the corresponding $\pi\pi$ two particle
energies. 
We also show the NS masses to illustrate the impact of the
disconnected correlator contribution $D(t)$. One notes that
the low-lying NS state does indeed track the $\pi\pi$ state at 
light quark masses~\cite{Prelovsek:2005du}.

With respect to the singlet meson operator fits, we note that
useful signals for $D(t)$ were obtained out to $t=7$, $9$ and
$10$ for ensembles  $A$, $B$ and $C$ respectively.
Stable fits with one scalar and one (fixed) oscillating state
were obtained with $t$ in the range $3$-$8$ or so. We were not able
to determine an excited scalar state.
Since glueball correlators on these data sets were
only well determined out to $t=3$ or $4$, the prospects
of performing fully factorising fits incorporating both glueball
and meson operators are not good with this level of statistics.

Unlike the NS case, the lowest energy singlet state with light
quarks (${\dgre \bullet}$)
lies  above or close to the effective two particle
threshold (${\blu\star}$). In this case the $\pi\pi$
decay channel is physically allowed.
In the flavour singlet case, the valence operator coupling to an $f_0$
state should include the strange quark.  
If included, this could well
raise the ground state further above this threshold.
Even with the correct strange quark content, the singlet state
({\dgre $\bullet$}) is likely to lie below the non-singlet
({\red $\triangle$}).

\section{Conclusions}
When improved staggered fermions are used to represent light dynamical quarks,
the glueball mass is not significantly different from the
quenched value at finite lattice spacing. 
Moderate taste violations are present in the scalar sector comparable
with those observed for the pion.
At light quark masses, decay channels complicate the mass determinations
as do the effects of taste breaking in intermediate states.
There is some evidence that the 2 flavour  singlet meson lies below the
corresponding non-singlet meson.
Further details may be found in~\cite{Miller:2005lu}.

\section{Acknowledgements}
We would like to thank R Edwards and B Joo
for assistance in code development with the SCIDAC-supported
code Chroma~\cite{Edwards:2004sx}, 
and Sasa Prelovsek for sharing her recent work on
staggered chiral perurbation theory.


\bibliographystyle{JHEP}
\bibliography{scalars}


\end{document}